# Deep learning–enabled image quality control in tomographic reconstruction: Robust optical diffraction tomography

DONGHUN RYU,[1,2] YOUNGJU JO,[1,2,3,5] JIHYEONG YOO,[3] TAEAN CHANG,[1,2] DAEWOONG AHN,[3] YOUNG SEO KIM,[2,3,4] GEON KIM,[1,2] HYUN-SEOK MIN,[3] AND YONGKEUN PARK[1,2,3,*]

[1]*Department of Physics, Korea Advanced Institute of Science and Technology (KAIST), Daejeon 34141, Republic of Korea*
[2]*KAIST Institute for Health Science and Technology, Daejeon 34141, Republic of Korea*
[3]*Tomocube Inc., Daejeon 34051, Republic of Korea*
[4]*Department of Chemical and Biomolecular Engineering, KAIST, Daejeon 34141, Republic of Korea*
[5]*Present address: Department of Applied Physics, Stanford University, Stanford, CA 94305, USA*
*\*yk.park@kaist.ac.kr*

**Abstract:** In tomographic reconstruction, the image quality of the reconstructed images can be significantly degraded by defects in the measured two-dimensional (2D) raw image data. Despite the importance of screening defective 2D images for robust tomographic reconstruction, manual inspection and rule-based automation suffer from low-throughput and insufficient accuracy, respectively. Here, we present deep learning–enabled quality control for holographic data to produce robust and high-throughput optical diffraction tomography (ODT). The key idea is to distill the knowledge of an expert into a deep convolutional neural network. We built an extensive database of optical field images with clean/noisy annotations, and then trained a binary-classification network based upon the data. The trained network outperformed visual inspection by non-expert users and a widely used rule-based algorithm, with > 90% test accuracy. Subsequently, we confirmed that the superior screening performance significantly improved the tomogram quality. To further confirm the trained model's performance and generalizability, we evaluated it on unseen biological cell data obtained with a setup that was not used to generate the training dataset. Lastly, we interpreted the trained model using various visualization techniques that provided the saliency map underlying each model inference.

## 1. Introduction

Quantitative phase imaging (QPI) has played a rapidly expanding role in biomedical applications, due to its label-free, quantitative live-cell imaging capability [1]. Combined with improved optical designs and computational algorithms, QPI can be expanded to advanced imaging modalities, such as synthetic aperture imaging [2], Fourier ptychography [3], and tomographic QPI [4-6]. Among them, optical diffraction tomography (ODT) is a three-dimensional (3D) QPI technique that reconstructs a 3D refractive-index (RI) tomogram, based on the Fourier diffraction theorem, from multiple 2D optical fields measured with various illuminations [4, 7]. Thus, the precise acquisition of the multiple 2D images is crucial for tomographic reconstruction. However, undesired noises can be added to the input fields, which deteriorate the image quality of the RI tomograms as they pass through the multiple stages of the ODT reconstruction pipeline (i.e., from hologram acquisition to field retrieval, along with phase wrapping to tomographic reconstruction).

A number of obstacles to robust tomographic reconstruction exist in practice. First, the spectral instability in laser perturbs the interference signals on holograms [8]. A small shift in the wavelength of the light source can result in substantial changes to the optical path length, thereby generating unintended interference patterns. Second, an illumination control unit, e.g., a digital micromirror device (DMD), may generate unwanted diffraction patterns on the holograms [6, 9]. Third, phase-unwrapping algorithms may fail to operate, thereby generating severe artifacts, including phase discontinuity [10]. Unexpected diffraction of the light passing through a sample, typically arising from dust particles or air bubbles inside the media, can result in abrupt changes in the detected hologram image. These high-gradient signals could cause the phase-unwrapping solver to obtain inaccurate solutions. To prevent possible degradation of the recovered tomogram quality, one can 'manually' exclude the noisy fields by eye, and utilize the remaining clean optical fields for the ODT reconstruction. To the best of our knowledge, no rigorous metrics for assessing the retrieved fields in ODT have been investigated. Furthermore, we believe it would be challenging to establish a rule-based metric that considers human perceptive assessment.

To build an effective framework for screening optical fields for robust tomographic reconstruction, considering the above, it is beneficial to employ data-driven approaches, e.g., machine learning and deep learning. With proper data preprocessing and annotation, these approaches can construct a powerful computational model for a wide variety of functions, by learning complex patterns from high dimensional data [11]. In particular, optical imaging, which deals with a tremendous number of pixels/voxels, has recently been taking full advantage of the data-driven approach [12]. Synergetic examples include image classification [13-17], segmentation [18, 19], resolution enhancement [20, 21], noise reduction [22, 23], light scattering [24-26], and *in-silico* labeling [27-29].

Here, we present a deep learning–based classification network that screens out noisy optical fields for improved tomographic reconstruction. Our network was trained using the classification standard of an ODT expert who annotated large datasets of individual bacteria. For verification, we compared our network with manual classification blinded to the training dataset on the test bacteria dataset, as the network aims for near-human-level classification performance. A rule-based classification algorithm, which computes the peak frequency signal of an optical field, was also compared to our method. To assess the generalizability of our method, we classified the optical fields of unseen NIH3T3 cells obtained from a different setup, and reconstructed a tomogram from the classified fields, which we then compared with the rule-based algorithm. Finally, we provided three different saliency maps that visualized the network's inference rationale, which could further validate our data-driven approach.

## 2. Methods

### 2.1 Optical-diffraction tomography and the screening of retrieved optical fields

The 3D RI tomogram of a sample was reconstructed from multiple 2D holograms, captured with variable-angle illuminations, based on the Fourier diffraction theorem [Figs. 1(A)−(B)] [4]. Mach-Zehnder interferometry was used for the hologram measurements, and a digital micromirror device (DMD) was utilized for systematic and rapid control of the illumination field [9, 30]. From the 71 measured holograms, corresponding optical fields were retrieved via a field-retrieval algorithm [31]. Then, optical fields having unwanted defects, due to the various sources explained above, were sorted out, to improve the image quality of the reconstructed tomogram. In Fig. 1(C), 54 clean and 17 noisy optical fields were manually screened. While the clean optical fields of *Escherichia coli* are clearly shown, the strong fringe patterns and broken optical fields degrade the image quality [Fig. 1(C)].

Two 3D RI maps were reconstructed from the two sets of input optical fields. The RI tomogram, which was reconstructed from all 71 measured fields [Fig. 1(D)], contains fringe patterns and ringing artifacts, which, in contrast, are not shown in the recovered tomogram that only uses the clean optical fields after the annotation [Fig. 1(E)]. Again, these visualizations signify that classifying retrieved optical fields and sorting them are essential in ODT reconstruction. Note that we did not use any regularization techniques, commonly utilized for addressing the missing-cone problem, to obtain an objective comparison of the image quality [32]. For the NIH3T3 test dataset used for the generalization test, we measured a tomogram of cells from a different ODT system, where the setup configuration was identical to the system used for bacterium imaging, except for the number of illumination angles (= 45), and the manufacturers of a few optical components. The details of the imaging system are explained in Appendix A.

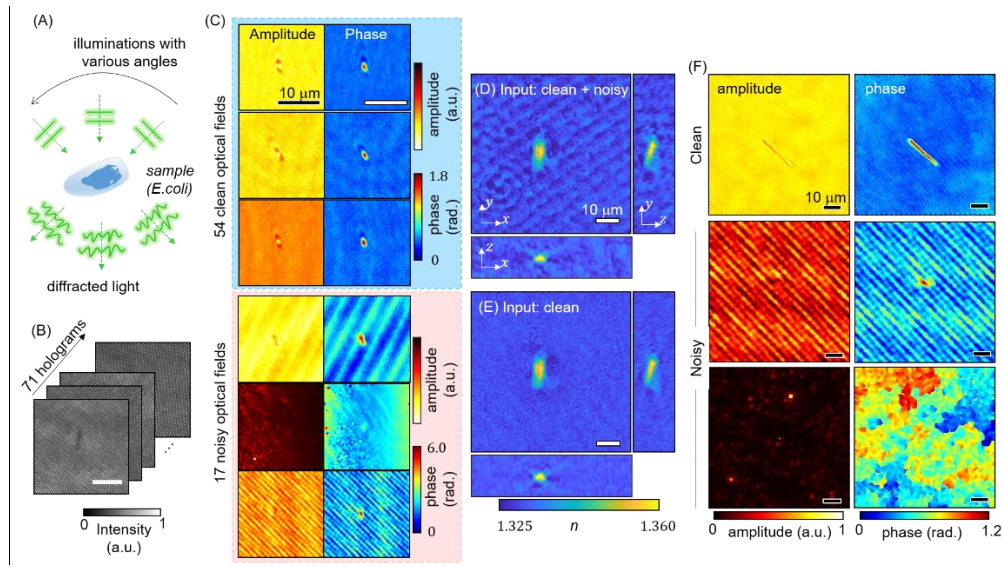

Fig. 1. (A) In ODT, a sample is illuminated from various angles. (B) Holograms obtained under various illumination angles. (C) Optical fields reconstructed from the holograms, including 54 clean and 17 noisy fields annotated by a trained ODT user. (D)–(E) RI tomograms of *E. coli*, recovered using all 71 fields and only the 54 clean fields, respectively. (F) Annotation standard. A trained individual has annotated the optical field data to prepare the training and test datasets. (Top) a bacteria cell is visible in the clean background. (Middle) strong unwanted fringe patterns overwhelm the sample, making it hard to identify. (Bottom) the optical fields are totally broken since unwanted noise was captured in the hologram intensity or no signals were obtained.

### 2.2 Data annotation

The total number of reconstructed optical fields of four species of bacterium (*E. coli*, *Klebsiella pneumoniae*, *Pseudomonas aeruginosa*, and *Staphylococcus epidermidis*) is 30,814. We imaged 434 RI tomograms; 71 optical fields were measured to reconstruct each tomogram.

Next, an ODT expert annotated the 30,814 complex fields, according to three types of optical field: one clean and two noisy, as depicted in Fig. 1(F). The recovered optical field was labeled as clean if a specimen was clearly visible and the background did not have significant defects. The first noisy-image case was corrupted with strong fringe patterns, emerging from unwanted interference signals. The second noisy-image case included broken images, due to the failure of the phase-unwrapping algorithm. This case typically occurred when the high-angle incident beam was diffracted by dust particles or bubbles in the sample medium, thereby blocking the sample signal at the detector. Both cases hindered the identification of the specimen of interest.

After balancing the two annotated datasets to have an equal number of images (= 2,660), we separated the obtained fields into 3,990 and 1,330 images, as training and test datasets, respectively. The NIH3T3 cells were annotated under the same standard. The test sets were never reviewed before the classification network was completely trained. The specific sample preparation is described in Appendix C.

*2.3 Classification-network implementation*

The convolutional neural network, designed for classifying input optical fields in the ODT reconstruction, is illustrated in Fig. 2. For training, we used a single-channel phase image ($1 \times 128 \times 128$) as an input to the network, because we observed that the phase image provides higher accuracy than a double-channel complex field ($2 \times 128 \times 128$) that concatenates the amplitude and phase, or a single-channel amplitude (See Appendix B).

The network architecture consists of two main functional blocks: feature extraction (C1–6) and classification (L1–3). The first block extracts various features using convolutional filters and non-linearity. The $1 \times 128 \times 128$ input images are processed to become 512 4×4-feature images, after passing through five consecutive sub-blocks consisting of a convolutional filter (the number of filters doubles in each sub-block (32, 64, 128, 256, and 512), rectified non-linearity (ReLu), and max pooling (slide = 2). Then, at the last C6, with the convolution, ReLu, and adaptive max pooling forcing it to have a single scalar output, 1,024 feature maps are generated before entering the fully connected layers for the classification.

The following three sub-blocks (L1−3), consisting of a dropout layer (p = 0.3, 0.5, and 0.5), a linear layer (1024-512-512-1), and a final sigmoid function (equation), shrink the 1,024 feature maps to a single scalar [0, 1]. This is used to infer whether the input image is defective or clean, given a threshold that can be tuned between 0 and 1. Here, we chose 0.5 for the threshold, based on the network performance.

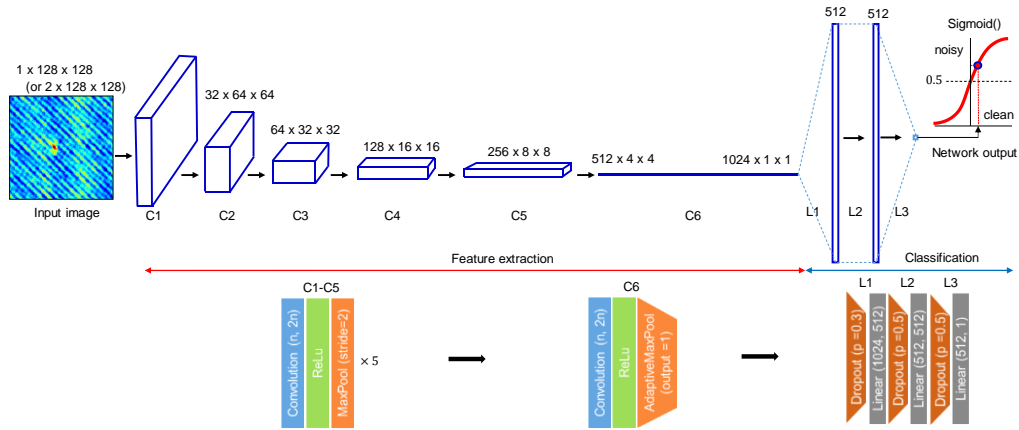

Fig. 2. The architecture of the deep neural network for classifying optical fields in ODT. The feature extraction part, composed of the convolution layer, ReLu (Rectified Linear Unit), max pooling, and adaptive max pooling, generates 1,024 feature maps. They are converted into a single scalar in the classification part for the final inference. C and L stand for convolutional block and linear block, respectively. n is the image dimension and p is the dropout rate.

The deep-learning model was implemented in a Pytorch 0.3.1 framework using a GPU server (two Intel® Xeon® CPUs E5-2620 v4, eight NVIDIA GeForce GTX 1080 Ti, and 11 GB memory). We center-cropped each image to $128 \times 128$ pixels, and carried out an elastic transform [33], which is widely used in data augmentation to accelerate network performance. We used an ADAM optimizer [34] and binary cross-entropy as the loss function to train the network. Typical training-epoch numbers range from five to 30 to achieve over 90% accuracy. We heuristically stopped the algorithm at epoch 20. For the single-channel phase dataset, the averaged training time for 10 different random initializations was approximately 20 minutes with a batch size of 16. We employed He's initialization [35], which considers the rectifiers as nonlinearities, for the convolutional layers, and Xavier's initialization [36] for the linear layers. We can classify an arbitrary input image in real time, once the network is trained.

## 2.4 Blind evaluation on a test dataset classified by human raters

Five physicists with various levels of ODT experience evaluated the pre-split test data of bacteria optical fields for comparison with our method. The test data were split into five sets, consisting of 200 images without repetition, and given to the raters. Note that the physicists were neither involved in the data acquisition nor exposed to the initial annotating process.

## 2.5 Rule-based algorithm for classifying optical fields

To benchmark the proposed classification network, we implemented a rule-based algorithm exploiting the peak signal of a given optical field in Fourier space. Because most biological samples are transparent (weakly scattered), low spatial-frequency components are dominant in the captured optical fields. By contrast, strong fringe patterns or abrupt phase changes could generate abnormally high values in the high spatial-frequency information (Fig. 3, left).

We also outline the procedure for the rule-based classification algorithm (Fig. 3, right). For each 2D optical field, we first compute the absolute spectrum (in $\log_{10}$) of the spatial Fourier transform. Second, the peak values outside a user-specified high-pass mask, centered at the illumination wave vector (i.e., sample low-spatial-frequency component), were used to screen the fields. To ensure that we focus on the high spatial-frequency region, the mask size was chosen to be large enough to occlude the low-spatial-frequency component. Here, we chose $40 \times (1 / 0.16)$ ($\mu m^{-1}$) as a mask diameter. Then, optical fields in which the peak value exceeded a user-specified threshold were classified to reconstruct a RI tomogram. For the screening threshold value, we heuristically chose 3.306 to screen the NIH3T3 optical fields.

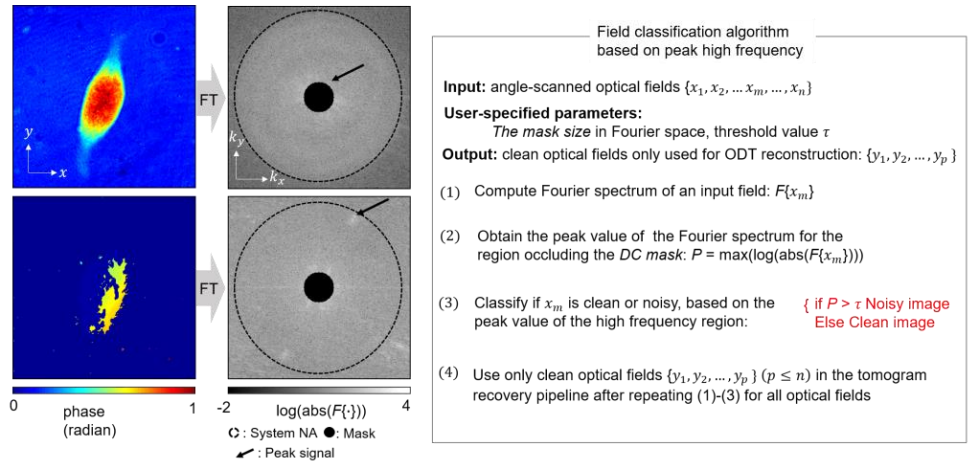

Fig. 3. Rule-based algorithm for classifying optical fields. (Left) clean and noisy NIH3T3 phase images and their Fourier spectrum. (Right) flowchart/procedure of the algorithm that computes the peak value of the high-frequency region for each optical field.

## 2.6 Network visualizations toward interpretable deep learning

Three visualization techniques were utilized to elucidate the spatial distributions of the image-activation map that led to our deep network's decision: class-activation mapping (CAM), guided back propagation (Guided BP), and gradient-weighted class-activation mapping (Grad-CAM), which all are described in more detail elsewhere [37-39]. All of these methods use localization maps, which conserve the spatial information of the network input in convolutional layers and expand them to an input-sized saliency map.

The CAM images illustrated below (see Results and Discussion) are generated at the last convolutional layer (C6 in Fig. 2) before entering the classification layers (L1−3). The guided BP images are generated at the end of C2. They are bilinearly upsampled to the dimension of the network input to clearly visualize the saliency-map distribution that our algorithm might have used for inference. Lastly, Grad-CAM images are created via the point-wise product CAM and guided BP products.

## 3. Results and Discussion

We first trained a neural network for the binary classification of optical fields using datasets consisting of four classes of bacteria (*E. coli, K. pneumoniae, P. aeruginosa,* and *S. epidermidis*). The convolutional neural network that determines whether a given optical field is clean or noisy was trained using gradient descent–based optimization and stopped with epoch 20, when the validation accuracy typically started to decrease in our algorithm (see Methods for details). Here, we present the averaged training and test accuracy for 10 different initializations of model weights. The training accuracy of the network achieved 98.97

± 0.68% (± standard deviation, SD). Specifically, the accuracies for the clean images (sensitivity) and noisy images (specificity) are 98.86 ± 0.97% and 99.09 ± 0.74%, respectively.

To test the performance of the trained network in terms of accuracy, we compared our proposed method with a classification that was screened by five human raters who had not seen the dataset, and a rule-based algorithm that computes a maximum value in the high-frequency region of Fourier space (Fig. 4, see also Methods). For the bacteria test data, the proposed method achieved 92.15 ± 0.07%, outperforming the averaged accuracy of the five human raters' classification, 76.22 ± 4.11%, and the rule-based algorithm, 65.71%, [Fig. 4(A)]. To study how much of each image class is correctly classified, the specificity (clean) and sensitivity (noisy) of the three methods are shown in Fig. 4(B). The proposed classification network obtained higher test accuracy: proposed method (92.28 ± 3.06% and 92.35 ± 2.57%), human raters (72.51 ± 9.19% and 79.93 ± 5.22%), and rule-based algorithm (95.94% and 35.49%).

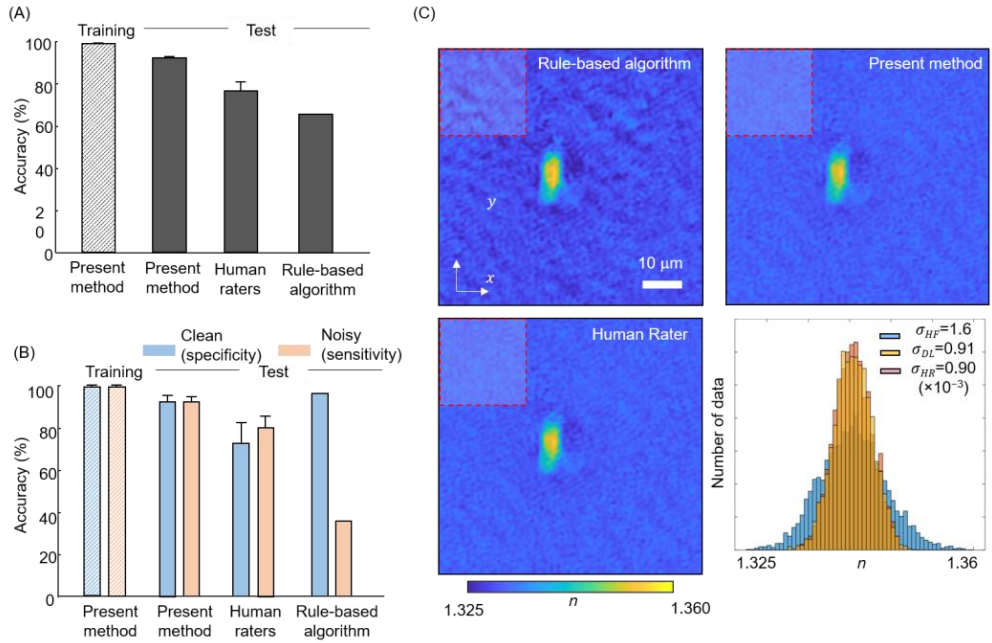

Fig. 4. Performance of deep learning, human raters, and rule-based algorithm on the field classification, and corresponding tomogram reconstructions. (A) Total training and test accuracy are compared for the present and three other methods. (B) For each case in (A), the specificity and sensitivity of two classes (clean and noisy) are investigated. (C) 2D sliced images of 3D tomograms are compared, and background noises are quantified via the standard deviation computed in the shaded region.

To investigate whether the surpassing test accuracy of the present method results in improved tomograms, we reconstructed three RI tomograms of *E. coli* from the three screened optical-field sets. The reconstructed images from the present method and the rater who achieved the best accuracy have fewer artifacts in the background, in comparison to the rule-based algorithm. We also measured the SD of the boxed region for the three reconstructed tomograms. The SDs of the present method, human raters, and rule-based algorithm are $0.91 \times 10^{-3}$%, $0.90 \times 10^{-3}$%, and $1.60 \times 10^{-3}$%, respectively [Fig. 4(C)].

Three remarks on this comparative study are noteworthy. First, while the test accuracies of the raters are approximately 16% lower than those of our method [Fig. 4(A)], their reconstructed image qualities and noise levels are not notably distinguishable. This seems to be because the raters and our classification network that learned expert perception are capable of sorting out noisy optical fields that would significantly deteriorate the reconstructed tomogram, which may not be screened by the naïve algorithm introduced here. Second, the standard deviations of the accuracies for the human raters are greater than those of the deep learning, which implies that the manual screening may not return a consistent output. Third, the image qualities (e.g., visibility of *E. coli*) of the reconstructed tomograms are not notably different, although the three optical-field datasets have a different number of angle illuminations. To some extent, the reduced number of illuminations in ODT can provide a reasonable quality of tomogram output, which has been more systematically investigated elsewhere [40, 41].

To further examine the proposed method and its generalizability, we imaged eukaryotic cells (NIH3T3) using an ODT imaging system that was not used to obtain the training set, as depicted in Fig. 5. Moreover, to consider an even worse situation, we chose the optical-field set that contained more noisy images than clean images for this test (22 clean and 23 noisy images). We first used 22 clean optical fields to reconstruct 3D tomograms, which were used as a gold standard to evaluate our method and the rule-based algorithm. Figures 5 (A)–(B) display reconstructed tomograms using all 45 optical fields, and using only the 22 clean fields as the ground truth, respectively. They visualize the significance of field screening in tomographic reconstruction.

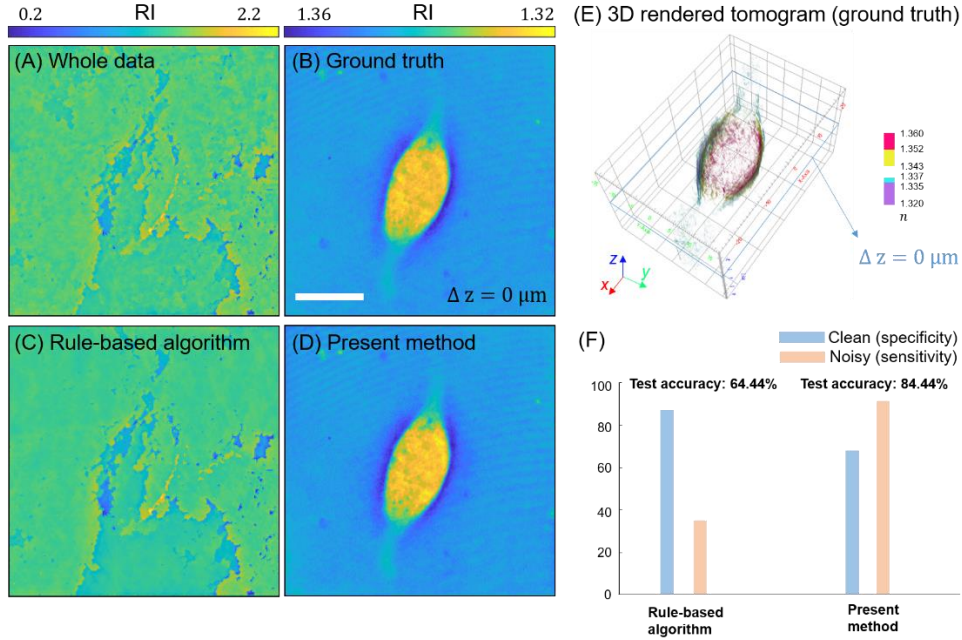

Fig. 5. Generalization test: NIH3T3 cell tomograms (sliced at axial center, Δ z = 0 μm) are recovered from various classified optical field data from (A) all 45 fields, (B) ground truth prepared by the expert, (C) the rule-based algorithm, and (D) the present method. (E) 3D iso-surface rendering of the tomogram recovered from the ground-truth optical fields. (F) The specificity and sensitivity, along with the total test accuracy, are depicted.

In Figs. 5 (C)−(D), our classification network generated a reconstructed tomogram with a similar quality level as the gold standard, in contrast to the poorly reconstructed tomogram using the rule-based algorithm. The total classification accuracy, specificity (clean), and sensitivity (noisy) of our method are 84.44%, 72.72%, and 95.65%, respectively; the accuracies of the rule-based algorithm for the same cases are 65.71%, 95.94%, and 35.49%, respectively. Including noisy fields in the reconstruction severely degrades the image quality of the RI tomogram, though the specificity (i.e., accuracy for clean fields) of the rule-based algorithm exceeds that of our method. This finding reemphasizes that screening optical fields in the ODT reconstruction is a critical process to determine the image quality of reconstructed tomograms.

To interpret the results, we obtained saliency maps that highlighted the spatial regions relevant to the network inference, using three widely used visualization techniques (see Methods). For the four types of image (clean/noisy *E. coli* and clean/noisy NIH3T3), the input phase image, corresponding class-activation mapping (CAM) images, guided back-propagation (GuidedBP) images, and gradient-weighted class-activation mapping (Grad-CAM) images are depicted in Fig. 6 (columns 1–4). The CAM images do not display the details of the interesting objects (i.e., *E. coli*, NIH3T3), but distinguish the objects from the background. Unlike the CAM images, the BP images provide global distribution of the activations with a higher resolution; however, the information around the specimens is less localized. Finally, the Grad-CAM images, inferred as clean (rows 1 and 3), elucidate a more localized detection of the target specimen, while parasitic fringe patterns and broken phases, dispersed across the images, are visible in the noisy images (rows 2 and 4). These saliency maps well represent the decision-making process of the algorithm, as one could look through the overall image and detect the object boundary or any defects in the image.

## 4. Conclusions and Outlooks

We proposed and experimentally demonstrated a deep neural network that screens optical fields to improve the image quality of RI tomograms in ODT. The proposed network, trained with an optical-field dataset annotated by a human expert, outperformed the overall classification results produced by human raters and the rule-based algorithm by a large margin (>20%). To understand the inner workings of our model beyond its results, we utilized various visualization techniques to obtain saliency maps that highlight the pixels relevant to the predictions. We envision that the proposed classification framework, replacing the perception of the well-trained experimentalist, would enable a fast, automated, and consistent tomographic reconstruction, which is possible for ODT and can be extended to tomographic-reconstruction methods in general.

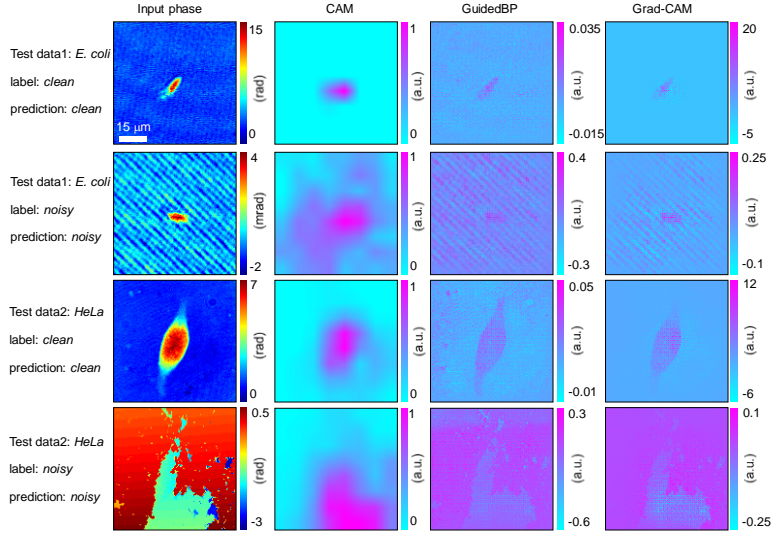

Fig. 6. Visual interpretations of the deep network inference. For each class (clean and noisy) of the phase image for *E. coli* and NIH3T3 (column 1), we extracted a class-activation mapping image (CAM) (column 2), guided back-propagation (GuidedBP) (column 3), and gradient-weighted class activation mapping (Grad-CAM) (column 4). Among the three visualization techniques, the Grad-CAM images, generated from the point-wise product of CAM and guided BP, best display the anatomical region that our deep neural network might have utilized for its inference.

Additional improvement aspects can be investigated to supplement our classification network. Firstly, our annotation standard can be further systematically established, rather than manually classifying input optical fields into three categories (clean, fringe noise and broken-image noise). We relied entirely on the experienced ODT expert's visual perception to make the qualitative ground-truth label for the dataset. Reconstructing a test sample with a known RI from various optical-field combinations may provide a quantitative method for selecting input fields to improve the reconstructed tomograms. Furthermore, a large number of annotators can make a better gold standard, which would enable the artificial intelligence (AI) model to learn from their collective intelligence.

Secondly, while only three visualization methods were utilized in this work to understand the network output, a number of interpretability methods are available for deep learning [42-44]. In addition, understanding already-trained models using such methods would facilitate the building of inherently interpretable deep-learning models. Thirdly, the proposed network used a narrow variety of training sets (four types of bacteria). Although the test set of NIH3T3 cells was successfully identified using our network, we expect that training with more diverse samples would generalize our deep-learning model.

Ultimately, though we built a classification network that should take *labeled* (i.e., *supervised*) input optical fields in this work, developing an *unsupervised* deep neural network for the same task remains as the next milestone toward a completely autonomous AI. For most supervised deep-learning techniques, manual annotation to prepare the labeled datasets should be conducted, which could be inconsistent and time-consuming. Computational tools that assist such laborious labeling or a trained lightweight network could be considered to improve the annotation. In this work, we employed one of the most widely used deep neural networks, a convolutional neural network, which has been successfully used for various problems in imaging science. Because it is technically difficult to obtain a ground-truth image in image-reconstruction problems, many existing deep learning–based image-regression models that statistically learn mapping functions between different types of images may not be feasible in general, unlike physical models capable of analytically explaining their principles. On the other hand, image detection and classification problems can establish a gold standard for most cases, which can be very reliable domains for deep-learning technology. We believe that various problems in imaging science, particularly those related to image classification, can continue to leverage AI technologies to aid or replace the tremendous amount of laborious work.

## Appendix A: Optical system of ODT

We used a commercial ODT system (HT-1S, Tomocube Inc.) with custom modifications [Fig. 7(A)]. The system is based on Mach-Zehnder interferometry. We used a diode-pumped solid-state laser beam (532 nm wavelength, 10 mW, MSL-S-532-10 mW, CNI laser, China) coupled into a 1×2 fiber coupler (OZoptics, Canada) that outputs a sample and a reference beam. The upward sample beam passing through Lens 2 is incident on a DMD (DLP6500FYE, Texas Instruments, USA) that generates many orders of diffraction beams, which was used to control the illumination angle. The first-order diffracted beam, passing through lens 3 and objective lens 1 (Numerical aperture (NA) = 0.7, ×60), impinges on a sample. Next, the scattered sample signal, conveyed by Objective lens 2 (×60, NA = 0.8) and the reference beam generate an off-axis hologram at the complementary metal-oxide-semiconductor camera (FL3-U3-13Y3M-C, FLIR Systems, USA). The captured holograms are utilized to reconstruct the 3D RI tomogram, according to the explained procedure in the main text.

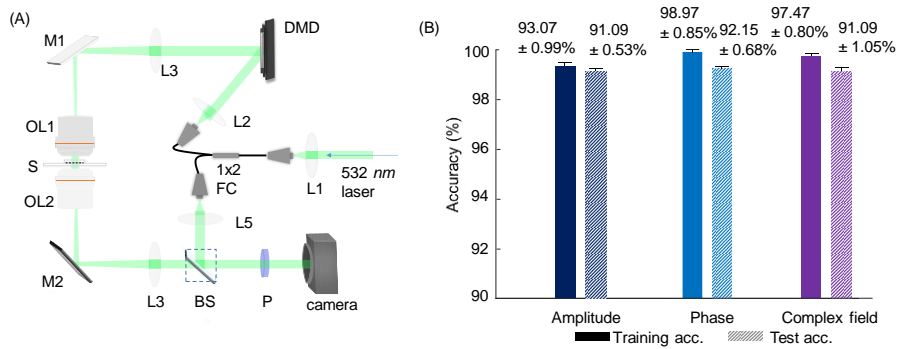

Fig. 7. (A) ODT system. L: Lens, FC: Fiber coupler, DMD: Digital micromirror device, M: Mirror, BS: Beam splitter, OL: Objective lens, S: Sample, and P: Linear polarizer. (B) Training and test accuracy of classification network using amplitude, phase, and complex field data.

The ODT imaging system, which was used to obtain NIH3T3 data, have several different optical components to the system that acquired the bacteria dataset; laser beam (632 nm wavelength, 15 mW, Thorlabs Inc., USA), an objective lens (NA = 1.1, LUMPLN, 60×, Olympus Inc., Japan), and an condenser lens (NA = 1.2 UPLSAPO, 60×, Olympus Inc., Japan).

## Appendix B: Training neural network using three types of input data

We decided to use phase images as input to our network, which achieved the best training and test accuracy in comparison to the results using amplitude and complex field concatenating amplitude and phase [Fig. 7(B)]. As for the described bacteria dataset, the training accuracy for amplitude, phase and complex field are $93.07 \pm 0.99\%$, $98.97 \pm 0.85\%$, and $97.47 \pm 0.80\%$; the test accuracy for the cases are $91.09 \pm 0.53\%$, $92.15 \pm 0.68\%$, and $91.09 \pm 1.05\%$. Though these accuracies rely on the complexity of network architecture, we anticipate that accuracy for the classification network trained using phase dataset would be higher than amplitude or complex field for most of the low-absorption samples.

## Appendix C: Sample preparation

The 3D RI tomograms of bacteria were acquired from specimens of *E.coli*, *K. pneumoniae*, *P. aeruginosa*, and *S. epidermidis* which were cultured in laboratory. The frozen glycerol stocks of bacteria, stored at -80°C, were thawed at room temperature. The bacteria were isolated from the glycerol solution by repetitions of centrifugation and washing with fresh media. After stabilization at 35°C incubator, the bacteria were streaked on agar plates. The agar plates were incubated in 35°C until colony formation was visible with naked eyes. Single colonies were seeded on fresh media using sterile loops. Each subculture was incubated at a 35°C shaking incubator until the concentration reached approximately 108-109 cfu/ml. The grown subculture was repeatedly centrifuged and washed, once with fresh medium and twice with phosphate buffered saline (PBS) solution. Finally, the washed bacteria solution was diluted with PBS solution until the concentration was adequate for single-specimen imaging.

The NIH3T3 cells (CCL-2, ATCC, Manassas, VA, USA) were preserved in Dulbecco's modified Eagle's medium (DMEM; Life Technologies) containing 100 U/mL streptomycin in humidified air (10 % $CO_2$), 10% (vol/vol) fetal bovine serum (FBS; Gibco, Gaithersburg, MD, USA), and 100 U/mL penicillin at 37°C.

## Funding

This work was supported by KAIST, Tomocube, and National Research Foundation of Korea (2015R1A32066550, 2017M3C1A3013923, 2018K000396).

## Acknowledgment

The authors thank Gunho Choi and Hyungjoo Cho (Tomocube) for helpful discussion on Grad-CAM and Pytorch implementation. The authors also thank Moosung Lee (KAIST) and JaeHwang Jung (Samsung electronics) for providing an initial test dataset. We are grateful to the KAIST Biomedical optics laboratory members for their generous help on the field evaluation. Y. Jo acknowledges support from KAIST Presidential Fellowship and Asan Foundation Biomedical Science Scholarship.